\documentclass[prb,footinbib,reprint,showpacs]{revtex4-1}%
\usepackage{epsfig}
\usepackage{amsfonts}
\usepackage{amsmath}
\usepackage{amssymb}
\usepackage{graphicx}%
\usepackage{natbib}
\begin{document}
\title{Theory of quantum dot spin-lasers}
\author{Rafa{\l} Oszwa\l dowski,$^{1,2}$ Christian G\o thgen$,^{1}$ 
and Igor \v{Z}uti\'{c}$^{1}$}
\affiliation{$^{1}$University at Buffalo,
State University of New York, Buffalo, NY 14260, USA}
\email{rmo4@buffalo.edu, zigor@buffalo.edu}
\affiliation{$^{2}$Instytut Fizyki, Uniwersytet Miko{\l }aja Kopernika,        
Grudzi\c{a}dzka 5/7, 87-100, Toru\'n, Poland}
\date{\today}

\begin{abstract}
We formulate a model of a semiconductor Quantum Dot laser with injection of
spin-polarized electrons. As compared to higher-dimensionality
structures, the Quantum-Dot-based active region is known to 
improve laser properties, including the spin-related ones. The 
wetting layer, from which carriers are captured into the active
region, acts as an intermediate level that strongly influences the lasing 
operation. The finite capture rate leads to an increase of lasing 
thresholds, and to saturation of emitted light at higher injection. 
In spite of these issues, the advantageous threshold reduction,
resulting from spin injection, can be preserved. The "spin-filtering"
effect, i.e., circularly polarized emission at even modest 
spin-polarization of injection, remains present as well. 
Our rate-equations description allows to obtain 
analytical results and provides transparent guidance for improvement
of spin-lasers.
\end{abstract}


\maketitle
\section{Introduction}\label{Sec.Intro}
Experiments on semiconductor spin-lasers have demonstrated the potential of 
spintronics to go
beyond the limits of devices relying solely on the carrier charge.
\cite{Zutic2004:RMP,Rudolph2005:APL,Holub2007:PRL,Hovel2008:APL}
These structures offer a practical path to realize
spintronic devices, which could be useful for communications and signal 
processing, rather than limited to magnetoresistive effects. 
Spin injection into lasers is implemented optically, 
when circularly polarized light imparts the photons' angular momentum to
the spin of carriers,\cite{Meier:1984,Fabian2008:SST,*Lu2009}
 or electrically, when a magnetic contact polarizes
carriers entering the semiconductor.\cite{Zutic2004:RMP}
Apart from the successful early demonstration of a spin-laser
based on a bulk-like layer of GaAs,\cite{Ando1998:APL} 
most experiments in this field
concentrated on structures with quantum well (QW) active regions, using
optical pump,
\cite{Hallstein1997:PRB,Rudolph2003:APL,Rudolph2005:APL,Fujino2009:APL} 
electrical injection 
\cite{Holub2007:PRL} or a combination of both.%
\cite{Hovel2008:APL,Gerhardt2010:SPIE}
Recently, however, an (In,Ga)As/GaAs quantum dot (QD) spin-laser with 
electrical injection 
has been demonstrated, lasing at temperatures 100 K higher than 
its QW counterparts.\cite{Basu2008:APL}
QDs close a succession of reduced-dimensionality structures: 
quantum wells and wires, which have replaced
bulk-like active regions of semiconductor lasers.%
\cite{Alferov2001:RMP,*Arakawa1982:APL}
They allow to control 
the number and spin of carriers, as well as the quantum-%
confinement geometry.
\cite{Abolfath2008:PRL,*Maximov2000:PRB} 
A quantum dot spin-laser combines the potential of spin-polarized injection with the
advantages of a QD-based active region,
such as low threshold, robust temperature performance, and narrow 
gain spectra.\cite{Asryan:2002,Sellers2004:EL}
In addition to these properties of conventional (spin-unpolarized) 
QD lasers, the long spin relaxation times,\cite{Fabian2007:APS}
characteristic for QDs, are advantageous for spin-lasers.

Spin-dependent effects in semiconductor lasers were studied at various
levels of com\-plex\-ity.%
\cite*{%
[{}]
[{ This seminal work included spin-related effects for
$P_{Jn}=0$, assuming that the (typically very different)
electron and hole spin
relaxation times are identical to each other. }]{SanMiguel1995:PRA}}
\cite{Dyson2003:JoOBQaSO,Vurgaftman2008:APL,Basu2009:PRL} 
A transparent rate-equations
(RE) approach to QW-based lasers has allowed to elucidate 
main consequences of the spin-polarized injection.\cite{Gothgen2008:APL}
An important finding of this QW model is that the injection threshold 
$J_{T}$, characterizing spin-unpolarized lasers, splits into two thresholds,
$J_{T1}<J_{T2}$, when the injected carriers are spin polarized.
When injection reaches $J_{T1}$ 
(majority threshold), the laser starts to emit photons
with one helicity (circular polarization), the other helicity joining at
$J_{T2}$, at which minority-spin electrons reach the threshold density. Both
experiments\cite{Rudolph2003:APL,Rudolph2005:APL,Holub2007:PRL} and 
theory\cite{Vurgaftman2008:APL,Gothgen2008:APL} have demonstrated an important
advantage of the spin-lasers over the unpolarized ones: $J_{T1}<J_{T}$,
assuming that all other parameters are identical. 
The threshold reduction, 
\begin{equation}%
r=1-J_{T1}/J_{T},
\label{eq:r}
\end{equation}%
would be largest
for fully spin-polarized electrons with infinite spin relaxation time, 
reaching as much as $r=5/9$.\cite{Gothgen2008:APL}
 According to the model, for any injection in the $J_{T1}$ to
$J_{T2}$ interval, the laser acts as a "spin-filter", i.e., it emits
 circularly polarized 
light, even if the spin polarization of injected carriers is small.
The relative width of this interval,%
\begin{equation}
d=\left(  J_{T2}-J_{T1}\right)  /J_{T},
\label{eq:dwidth}
\end{equation}%
increases with the injected spin polarization.
The "filtering" effect is another merit of spin-lasers, as it offers new
opportunities for their dynamic operation. Modulation of injected 
spin polarization was shown to modulate the intensity of laser emission, 
even at a constant total injection, and to increase the modulation bandwidth.%
\cite{Lee2010:P} 

So far, theoretical description of spin-lasers has been essentially
 limited to QW-based models.
 To find distinguishing features of QD spin-lasers, in this work we 
formulate a model,
 which allows for analytical results and offers a direct comparison
 with the previous results for the QW spin-lasers.
\cite{Rudolph2003:APL,Rudolph2005:APL,%
Holub2007:PRL,Gothgen2008:APL,Lee2010:P} Here, we focus on the
 parameters motivated by the experiments on (In,Ga)As-based QD spin-lasers,%
\cite{Basu2008:APL,Basu2009:PRL},%
\cite*{%
[{Spin-dependent properties of (In,Ga)As QDs have been extensively studied,
 e.g. see, }]%
[{. (In,Ga)As has been frequently used for conventional lasers with QW
or QD active regions, see e.g., I. Tangring, H. Q. Ni, B. P. Wu, D. H. Wu,
Y. H. Xiong, S. S. Huang, Z. C. Niu, S. M. Wang, Z. H. Lai, and A. Larsson,
Appl. Phys. Lett. {\bf 91}, 221101 (2007).}]
{Yakovlev:2008}}. It is, however, instructive to consider 
other possible materials for spin QW and QD lasers, since a variety of active
 regions has been used for their conventional (spin-unpolarized) counterparts. 
This choice can be guided by long spin relaxation time for electrons,
 which enhances the desirable spin-laser characteristics.%
\cite{Oestreich2005:SM} Longer spin relaxation times can result,
 for example, from a reduction of spin-orbit coupling, one of the main
 sources of spin relaxation.\cite{Zutic2004:RMP,Fabian2007:APS} This can
 be achieved by choosing materials with light elements or by using different
 growth orientation in QWs.%
\cite*{[{A more than ten-fold increase in spin %
 relaxation time is possible by replacing (001) by (110) GaAs-based QW: }]%
{Fujino2009:APL}}
Long spin relaxation times have been reported in
 CdSe/ZnSe (an example of a II-VI structure)
self-organized QDs.\cite{Klochikhin2008:SST} Detailed predictive
 studies of the spin relaxation mechanisms in QDs%
\cite{Fabian2007:APS,Stano2006:PRB} will serve an important role in
 future efforts in designing QD spin-lasers.
 It would also be interesting to consider active regions with magnetic doping,
 where the spin degeneracy of the lasing transition may be lifted. II-VI
 materials doped with Mn are a promising direction, since QD lasers based on 
II-VI structures have already been considered.\cite{ Passow2002:ASSP} The problem of the 
Mn
 internal transition, which reduces the intensity of band-to-band transitions
 can be addressed by using ZnSe/(Zn,Mn)Te epitaxial QDs,\cite{Sellers2009:P}
characterized by a relatively low fundamental transition energy. 

A very interesing
 emerging field are  lasers based on colloidal semiconductor QDs [typically
 II-VI, such as CdS, CdSe, ZnSe, and ZnTe.%
\cite{Klimov2007:ARPC,Scholes2008:AFM}]
 These nanostructures are easily synthesized, offer a large tunability of
 transition energies and  a long spin-coherence time.\cite{Stern2005:PRB} 
Some colloidal QD structures suffer, however, from the very fast $(<100$ ps)
 non-radiative Auger recombination that hinders population inversion and is 
therefore 
detrimental for optical gain. This effect can be avoided by using the so-called
 type-II band alignment, in which spatial separation of electrons from
 holes significantly suppresses the Auger recombination.\cite{Klimov2007:N}
 Just like their self-assembled counterparts, collloidal QDs can be doped
 magnetically.\cite{Beaulac2008:AFM}

\section{Rate-Equations Model}
\begin{figure}[h]
\begin{center}
\includegraphics%
[angle=-90, scale=0.50,clip=true, viewport=1.2in 0.5in 7.7in 7.2in]%
{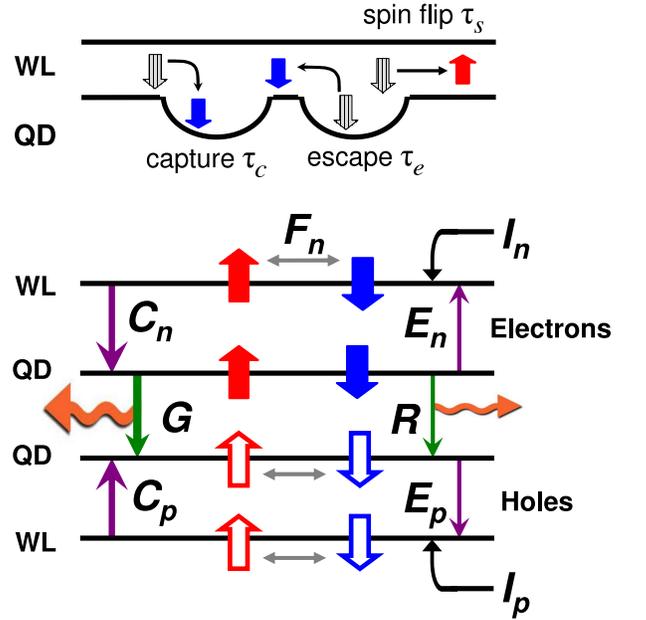}
\end{center}
\caption{(Color online) 
Processes 
in our model of a spin-laser, described by 
Eqs.~(\ref{eq:fw}-\ref{eq:G}).
QD: quantum dot, WL: wetting layer.
Upper panel: thick arrows denote electron spin direction in 
processes labeled by their corresponding times. Lower panel: 
thick vertical arrows show the carrier spin (filled for electrons, empty for 
holes). 
Curved arrows show carrier injection {\it I}. 
Thin arrows depict  capture {\it C}, escape {\it E}, 
spin relaxation {\it F}, stimulated ({\it G}) and spontaneous ({\it R})
recombination (thickness indicates 
relative rates). The subscripts $n$ and $p$ represent the electron
and hole contributions, respectively. 
Wavy arrows depict photon emission.
}
\label{fig:QDschematic}%
\end{figure}

The cavity of the QD spin-laser is in resonance
with interband transitions between QD-confined levels.\cite{Basu2008:APL}
Since the levels are
derived from valence and conduction bands, a general description requires
keeping track of both electron and hole populations, as previously shown
both for bipolar spintronic devices,\cite{Zutic2003:APL,*Zutic2006:PRL,*Zutic2007:JPCM} and 
for QD spin-unpolarized lasers.\cite{Fiore2007:IEEEJQE} 
The QDs capture electrons and holes from energy levels of a two-dimensional 
QW-like wetting layer (WL), 
which acts as a reservoir of carriers.
\cite{Dery2005:IJoQE},
\cite*{%
[{Treatment of the WL as a single level is justified by the 
relatively fast energy-relaxation processes to the lowest level 
(respectively for electrons and holes) in the wetting layer, see }]{Dery2004:IJoQE}}
Figure~\ref{fig:QDschematic} depicts the level structure and the 
various processes represented by our REs, from carrier injection 
to photon emission.
We describe the carriers by eight spin-resolved REs,
coupled to two REs for two circular polarizations of stimulated emission:%
\begin{eqnarray}
df_{w\alpha\pm}/dt&=&I_{\alpha\pm}-C_{\alpha\pm}+\frac{2}{\kappa_{\alpha
}}  E_{\alpha\pm}-R_{w\pm}\mp F_{w\alpha},  \label{eq:fw} \\%
df_{q\alpha\pm}/dt&=&\frac{\kappa_{\alpha}}{2}  C_{\alpha\pm}%
-E_{\alpha\pm}-R_{q\pm}-G_{\pm}\mp F_{q\alpha},  \label{eq:fq} \\%
df_{\mathrm{S}\mp}/dt&=&\Gamma G_{\pm}-f_{\mathrm{S}\mp}/\tau_{\mathrm{ph}%
},\label{eq:fS}%
\end{eqnarray}
cf. Fig.~\ref{fig:QDschematic}.
The index $w$ stands for WL and $q$ for QDs, while $\alpha=n,$ $p$
for electrons and holes, respectively. Equations~(\ref{eq:fw}) and 
(\ref{eq:fq}) describe carrier occupancies, $0\le f\le 1$, 
in WL and QDs, related to the corresponding 
numbers of particles
$n_{w\alpha}$ and $n_{q\alpha}$: 
\begin{eqnarray}
f_{w\alpha\pm} &=& n_{w\alpha\pm}/\left(N_{w\alpha}/2\right), \\
\label{eq:fw2}
f_{q\alpha\pm}   &=& n_{q\alpha\pm}/N_q.
\label{eq:fq2}
\end{eqnarray}
Here $N_{w\alpha}$ is the number of states in WL and
$N_q$ is the number of QDs. The ratio
$\kappa_{\alpha}=N_{w\alpha}/N_q$, used in Eqs.~(\ref{eq:fw}) 
and (\ref{eq:fq}), is an important parameter of the
QD laser.\cite{Summers2007:JAP}
For simplicity, we assume
that each QD hosts one double-degenerate level per
species $\alpha$. 
This can be realized only for electron levels in small enough QDs, 
\cite{Grundmann1996:PRB}
but we do not expect our results to be
qualitatively changed upon inclusion of QD excited states. As 
long as the lasing transitions involve only QD-confined levels,
the limited density of QD states and the limited capture rate will affect the 
spin-laser characteristics in the way discussed below.
The ground state of holes is assumed to be formed predominantly
from heavy-hole wavefunctions.
The electron (hole) level is degenerate with respect to
spin $\pm1/2$ (angular momentum $\pm 3/2$) projection.\cite{Zutic2004:RMP}

Equations~(\ref{eq:fS}) is for
photon occupancies, $f_\mathrm{S}$, of helicities $\mp$, defined as
\begin{equation}
f_\mathrm{S\pm}=S^{\pm}/N_q,
\end{equation}
where $S^\pm$ is the number of cavity photons of the given helicity.
Our
sign convention for indices denoting the spin projections and helicities 
follows Ref.~\onlinecite{Gothgen2008:APL}. 
In Eq.~(\ref{eq:fS}), $\Gamma$ is the optical
confinement factor and $\tau_{\mathrm{ph}}$ is the photon cavity
lifetime. The terms%
\begin{eqnarray}
I_{\alpha\pm}&=&J_{\alpha\pm}%
\left(  1-f_{w\alpha\pm}\right),
\label{eq:I} \\
C_{\alpha\pm}&=&f_{w\alpha\pm}\left( 1-f_{q\alpha\pm}\right)/\tau_{c\alpha}, 
\label{eq:C} \\
E_{\alpha\pm}&=&f_{q\alpha\pm}\left(  1-f_{w\alpha\pm}\right)/\tau_{e\alpha}, 
\label{eq:E}
\end{eqnarray}
represent carrier injection,  
carrier capture from the WL to QDs, and the
inverse process of escape, respectively.
$J_{\alpha\pm}=\left(  1\pm P_{J\alpha}\right)\! J_{\alpha}$ 
is the number of carriers of $\alpha$ species 
injected into the laser per WL state of the given spin and unit time,
with $J_\alpha=\left( J_{\alpha+}+J_{\alpha-}\right)/2$.
The injection spin polarization 
is $P_{J\alpha}=\left(J_{\alpha+} -J_{\alpha-}\right)%
/ \left(J_{\alpha+} +J_{\alpha-}\right)$.
The parameters $\tau_{c\alpha}$ and $\tau_{e\alpha}$ are the capture
 and escape times. 

To correctly describe consequences of the small density of QD states,
as well as saturation of the WL states at high injection,
it is important to include in 
Eqs.~(\ref{eq:I})--(\ref{eq:E})
the Pauli-blocking factors, $\left(  1-f\right)$, of
the WL and QD states.\cite{Summers2007:JAP} 
These terms, omitted in some previous work on QD-based spin-lasers,
\cite{Basu2009:PRL} impede carrier transfer to states close to saturation.
We find that they are particularly important in description
of the limited QD occupancies, as shown below.

Defining $\gamma=w,q$, we 
write the spontaneous radiative recombination 
in Eqs.~(\ref{eq:fw}) and (\ref{eq:fq}) as
\begin{equation}
R_{\gamma\pm}=b_{\gamma}f_{\gamma n\pm}f_{\gamma p\pm}, 
\label{eq:R}
\end{equation}
where $b_{\gamma}$ gives the recombination rate.
The spin-relaxation terms 
\begin{equation}
F_{\gamma\alpha}=\left(  f_{\gamma\alpha+}-f_{\gamma\alpha-}\right)  
/\tau_{s\alpha\gamma},
\label{eq:F}
\end{equation}
equilibrate spin sub\-po\-pulations with relaxation times
$\tau_{s\alpha\gamma}$.
A realistic model of a steady-state or dynamic operation of spin-lasers, 
should reflect the different behaviors of electron and hole spins. 
\cite{Gothgen2008:APL,Lee2010:P} %
Due to the strong spin-orbit coupling in the valence band, 
the spin polarization of holes is lost relatively 
quickly, i.e., $\tau_{sp\gamma}\ll\tau_{sn\gamma}$, both in
QWs (i.e., also in the WL) and QDs.%
\cite{Zutic2004:RMP,Rudolph2003:APL,Hall2007:APL}
Therefore we  assume that the holes, unlike electrons, are spin-unpolarized,
 i.e.,
$P_{Jp}=0$ and $f_{\gamma p\pm}=f_{\gamma p}$, which implies $I_{p+}=I_{p-}$
in Eq.~(\ref{eq:I}).
Additionally, the electron spin-relaxation in QWs is faster than in
QDs,
thus we take $\tau_{snq}\rightarrow\infty.$ This a very good approximation
at low temperatures,\cite{Paillard2001:PRL} and it remains reasonable
at room temperature, where $\tau_{snq}$ reaches 1 ns.\cite{Robb2007:APL}

The gain term in Eqs.~(\ref{eq:fq}) and (\ref{eq:fS}) 
\begin{equation}
G_{\pm}=g\left(  f_{qn\pm}+f_{qp\pm}-1\right)  f_{\mathrm{S}\mp}, 
\label{eq:G}
\end{equation}
describes coupling of the carriers and light, which gives rise
to stimulated emission. The sign ordering in subscripts is consistent
with the optical selection rules for interband transitions.\cite{Zutic2004:RMP}
The constant $g$ is independent of photon occupancies $f_{\mathrm{S}\pm}$,
i.e., it does not contain the gain compression terms.%
\cite{Carroll:1998,Chuang:2009} %
In spite of that, our QD model naturally predicts light-output 
saturation due to the limited capture capacity of QDs, as discussed below. 
We note that, owing to the above-mentioned spin asymmetry between electrons
and holes, the assumption $f_{qn\pm}=f_{qp\pm}$ is not justified for
$P_{Jn}\neq0$.
Thus, an attempt to express, e.g., $G_+$ [Eq.~(\ref{eq:G})]
using only $f_{qn+}$ (and $f_\mathrm{S-}$),
 would lead to incorrect threshold values, even for the QW spin-laser model.
\section{Results}
We focus on the steady-state regime, in which the total charge in the 
spin-laser is constant.
This imposes a relation between $J_p=J_{p+}=J_{p-}$ and $J_{n\pm}$.
One of the REs for carriers then becomes linearly dependent on the others,
and we replace it with the condition of overall charge neutrality.
In the spirit of the simple RE approach, we neglect 
carrier-carrier Coulomb interactions, which may become important 
at high injection.\cite{Schneider2001:PRB}

We have obtained all formulas presented below by solving the REs analytically.
To give simple expressions that offer insight into the behavior of the
spin-laser, we assume $\Gamma=1$, $\kappa_{\alpha}=\kappa,$ $R_{w\pm}=0,$
$\tau_{c\alpha}=\tau_{c}$, and $\tau_{e\alpha}=\tau_{e}.$
\cite*{[{
Consequences of $\tau_{cn}/\tau_{en}\neq\tau_{cp}/\tau_{ep}$ for
spin-unpolarized QD lasers have been discussed in }]{Viktorov2005:APL}} 
We have checked that the spontaneous-emission
coupling to the lasing mode has a negligible effect on our 
results.\cite{Gothgen2008:APL,Lee2010:P} Thus we
set the coupling factor $\beta=0$. This allows for an unambiguous
determination of the laser thresholds.

To develop a preliminary understanding of the QD model of a spin-laser, 
we relate it to the simpler QW model, discussed
in Sec.~\ref{Sec.Intro}. 
In the limit of $\tau_{c}\rightarrow0$ and $\tau_{e}\rightarrow\infty$,
$f_{w\alpha\pm}$ vanish, as can be inferred from Fig.~\ref{fig:QDschematic}. 
In this case, WL plays no role in 
the above QD model, which becomes "QW-like", i.e., similar (but not identical)
to the QW model of Sec.~\ref{Sec.Intro}.  
We emphasize that it is not our goal here to compare the absolute thresholds
 of a QW- and a QD-based laser. Such a comparison requires distinct
parameters for these two structures, and shows the potential for
achieving lower thresholds in the latter.
\cite{Asryan:2002,Bimberg:1999,Blood2009:STiQEIJo}
Here, we use the same range of parameters for the QD model
and for its QW-like limit (except for $\tau_{c,e}$). Thus, the QW-like
 model leads to \emph{lower} thresholds, since it describes
effectively a QD-based structure in the limit of instant capture.
Nevertheless, this approach enables us to 
elucidate important qualitative differences between the QW- and QD-based 
spin-lasers.  

\begin{figure}[tbh]
\includegraphics[scale=0.65]{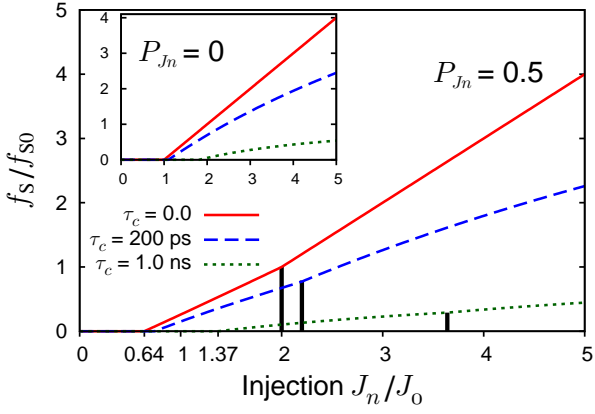}%
\caption{(Color online)
Main panel: 
Dependence of QD spin-laser emission on electron injection,
shown for different capture times $\tau_c$.
Total photon occupancy, 
$f_\mathrm{S}=\left(f_\mathrm{S+}+f_\mathrm{S-}\right)\!/2$,
is normalized to $f_\mathrm{S0}$ [Eq.~(\ref{eq:fS0})], 
while the electron injection to
$J_0$ [Eq.~(\ref{eq:J0})].
The parameters are $\tau_\mathrm{ph}=1$ ps, $b_q\tau_\mathrm{ph}=0.01$, 
$g\tau_\mathrm{ph}=2$, $\kappa=100$, $\tau_{snw,e}\rightarrow\infty$,
and $P_{Jn}=0.5$.
Vertical lines denote minority thresholds $J_{T2}$. The smallest ($\tau_c=0$)
majority threshold $J_{T1}$ for $P_{Jn}=0.5$, marked at 0.64, gives 
the threshold reduction $r$=0.36 [Eq.~(\ref{eq:r})]. Inset:
Results for spin-unpolarized lasers, $P_{Jn}=0$,
with the other parameters' values same as in the main panel.}%
\label{fig:tauc}%
\end{figure}

First, we consider consequences of the finite capture rate, 
$\tau_c>0$, for a spin-unpolarized laser, $P_{Jn}=0$, illustrated in
the inset of Fig.~\ref{fig:tauc}. 
Let $J_T$ be the threshold for a given $\tau_c$.
For any $J_{n}>J_{T}$, the QD occupancies are independent of $\tau_{c}$ and
fulfill $f_{qn\pm}=f_{qp}=f_{0}$,
where $f_{0}$ is the occupancy pinned at
the threshold value,\cite{Summers2007:JAP}
\begin{equation}
f_{0}=1/2+1/\left(  2g\tau_\mathrm{ph}\right).%
\label{eq:f0}%
\end{equation}
We normalize the light-injection characteristics using quantities 
in the limit of instant capture, $\tau_c=0$. The total 
photon occupancy, 
$f_\mathrm{S}=\left( f_{\mathrm{S}+}+f_{\mathrm{S}-}\right)\!/2$,
 is expressed in terms of
\begin{equation} 
f_\mathrm{S0}=f_\mathrm{S}(\tau_c=0,P_{Jn}=0,J_n=2J_T)=b_q 
\tau_\mathrm{ph} f_0^2,
\label{eq:fS0}
\end{equation}
while the injection $J_n$ is normalized to
\begin{equation}
J_{0}= J_{T}\left(  \tau_{c}=0\right)=2b_{q}f_{0}^{2}/\kappa.
\label{eq:J0}
\end{equation}
Unlike the pinned occupancies, $J_{T}$
increases with $\tau_{c}$ (Fig.~\ref{fig:tauc}, 
inset) as%
\begin{equation}
J_{T}=\left[  1+\frac{2f_0}{\kappa \left(1-f_0\right)}
\frac{\tau_c}{\tau_e}\right]\frac{J_L J_0}{J_L-J_0}, 
\label{eq:JT}
\end{equation}
where 
$J_{L}=\left(  1-f_{0}\right)  /\tau_{c}$ 
is the maximum capture rate $C$,
Eq.~(\ref{eq:C}), realized for $f_{wn}=f_{wp}=1$. 
The factor $(J_L-J_0)$ in Eq.~(\ref{eq:JT}) imposes an upper
limit on $\tau_{c},$ 
above which lasing is impossible $\left(  J_{T}%
\rightarrow\infty\right)  $. The limiting condition, $J_{L}\geq J_{0},$ means
that $J_{L}$ must overcome the recombination 
losses, $b_{q}$, determining $J_{0}$.
When $\tau_c \rightarrow 0$, the threshold $J_T$ 
reduces to $J_0$ from Eq.~(\ref{eq:J0}).

In a model of a QW laser with no gain compression, 
the laser light intensity depends linearly on 
injection (we neglect the small deviations from linearity that
appear around the thresholds when the coupling factor $\beta>0$).
A linear dependence is also found for the QD model with $\tau_c=0$. 
In contrast,
the QD model with $\tau_c>0$ reveals a 
sub-linear dependence
(Fig.~\ref{fig:tauc}, inset), even though we do not introduce
any gain-compression terms. 
\cite*{%
[{}]
[{ reports findings for $P_{Jn}=0$ with additional levels and 
introducing additional 
gain compression factor $\epsilon$, defined in Ref.\ 45.}]{Sugawara1997:APL}}
At higher injection
the emission saturates, as discussed
below for the spin-polarized injection scenario.

Next, we turn to the spin-polarized injection, i.e., 
$P_{Jn}\neq0$. Similarly to the QW model from Sec.~\ref{Sec.Intro}, 
our QD model predicts two lasing thresholds,
\footnote{This is true for parameters that give a finite 
$J_T$ [see Eq.~(\ref{eq:JT})], and
except such special cases as $P_{Jn}=\left| 1\right|$,
 $\tau_{snw}\rightarrow\infty$, in which $J_{T2}$ is never   
reached, (i.e., one of $f_{\mathrm{S}\pm}$ is zero for any $J_{n}$)%
.} 
$J_{T1}<J_{T2}$, as shown in Fig.~\ref{fig:tauc}, main panel. 
We find that, in general,
the increase of $J_{T1}$ and $J_{T2}$ with $\tau_{c}$ 
is quantitatively similar to the increase of $J_{T}$. 
A particularly simple example is the minority threshold in the limit of
$\tau_{snw}\rightarrow\infty$  :
\begin{equation}
J_{T2}/J_{T}=1/\left(  1-\left\vert P_{Jn}\right\vert \right)
\label{eq:JT2}
\end{equation}
valid for any $\tau_c,\ \tau_e,\ b_q,\ g,\ 
\tau_\mathrm{ph}$, $\kappa$, and identical to the 
relation found for the QW-based laser.

Such simple, universal relations are typical for the QW model, 
but not for the QD one with $\tau_c>0$.
Even with the simplifying assumptions:
 $b_{q}\tau_{c}\ll1$, $g\tau_\mathrm{ph}=2$, 
large $\kappa$ (i.e., $f_{wn\pm},f_{wp}\ll1$),
and $\tau_{snw,e}\rightarrow\infty$, we obtain 
a more complicated ratio for the majority threshold
\begin{eqnarray}
\frac{J_{T1}}{J_T}&=&\frac{4}{\left( 2+\left\vert P_{Jn}\right\vert \right)^2}
\nonumber \\
&\times& \left[ 1 + \frac{18 \left| P_{Jn}^{3}\right|b_{q}\tau _{c}}
{1+ 6\left| P_{Jn}\right| +3 P_{Jn}^2 -10\left|P_{Jn}^3\right| }\right],
\label{eq:JT1}
\end{eqnarray}
showing that the threshold reduction $r$,  Eq.~(\ref{eq:r}), 
depends on $\tau_c$.
Equation~(\ref{eq:JT1}) reduces to the simple QW-model result
 for $\tau_c= 0$ (Eq.~4 of Ref.~\onlinecite{Gothgen2008:APL} 
in the limit  w$\rightarrow 0$,
i.e., infinite spin relaxation time) .

\begin{figure}[tbh]
\begin{center}
\includegraphics[scale=0.85]{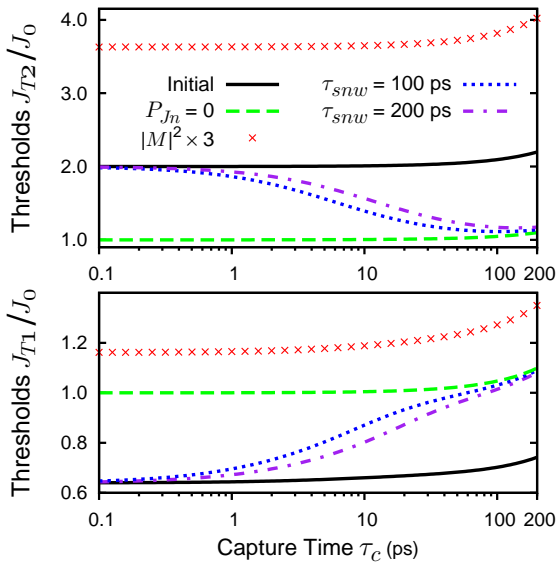}%
\end{center}
\caption{(Color online)
Majority and minority (lower and upper panel) thresholds of a QD
spin-polarized laser. The solid line shows our result for the initial 
parameters given in Fig.~\ref{fig:tauc}. The crosses,
dash-dot, and dotted lines are the thresholds when one of the parameters 
is changed (see legend). A three-fold increase of the squared modulus 
of the optical matrix element, $\left|M\right|^2$, results in a three-fold 
increase of both $b_q$ and $g$.
Dashed line in both panels is $J_{T}$ for $P_{Jn}=0$. 
The normalizing current $J_{0}=J_{T}\!\left(  \tau_{c}=0\right)$, 
Eq.~(\ref{eq:J0}), has been calculated for fixed parameters (the initial
parameters of Fig.~\ref{fig:tauc}, except $P_{Jn}=0$).}%
\label{fig:thresh}%
\end{figure}

Figure \ref{fig:thresh} shows the evolution of $J_{T1}$ and $J_{T2}$ 
as a function of the capture time. We use a range of $\tau_c$,
which reflects the scope of values found in previous works.%
\cite{Fiore2007:IEEEJQE,Giorgi2001:APL}
 We start from an initial set of parameters: $\tau_\mathrm{ph}=1$ ps,
\cite{Rudolph2005:APL}
 $b_{q} \tau_\mathrm{ph}=0.01$,
\cite*{%
[{Typical $b_q \tau_\mathrm{ph}$ values are $\sim 10^{-3}$, }]
[{. We use a larger value ($10^{-2}$), to effectively take into account the
recombination in the WL, since we set $R_w=0$ for transparency
of our approach. Losses outside QDs may be important for laser modeling
 (Ref.\ 50).
}]{Cao2004:APL}}
$g\tau_\mathrm{ph}=2$,\cite{Melnik2006:OE} 
$\kappa=100$,\cite{Matthews2002:APL} 
$P_{Jn}=0.5$, $\tau_{snw,e}\rightarrow\infty$,  and
then we vary some of the values to determine the relevant trends. 
The limit $\tau_{snw}\rightarrow\infty$ enables us to obtain analytical
formulas [such as Eq.~(\ref{eq:JT1})], 
we also present numerical results for $\tau_{snw}=$ 100 and 200 ps,
i.e., the order of magnitude found in experiments.
\cite{Jarasiunas2004:SST,*Schreiber2007:PSSB}
Both $J_{T1}$ and $J_{T2}$ increase with $\tau_{c},$ since the
capture rate into the QDs, Eq.~(\ref{eq:C}), decreases. Comparing $J_{T1}$ 
to $J_{T}$
(solid and dashed line, lower panel), we note only a small decrease in the
threshold reduction defined in Eq.~(\ref{eq:r}); 
$r\!\left(  \tau_{c}=200~\mathrm{ps}\right)  =0.32$,
versus $r\!\left(  \tau_{c}=0\right)  =0.36$. Using these values,
we calculate the "spin-filtering" interval, Eq.~(\ref{eq:dwidth}), from 
Eqs.~(\ref{eq:r}) and (\ref{eq:JT2})
for $P_{Jn}=0.5$. It decreases monotonically from the maximum 
$d=1.36$ for $\tau_{c}=0$
to $d=1.32$ for $\tau_{c}=200$ ps, only a small shrinking
of the "filtering" region. 

In the limit of $\tau_{snw}\rightarrow\infty$, we find that 
$J_{T1}$, $J_{T2}$, and $J_{T}$
rise uniformly with decreasing capture rate for a 
wide range of parameters, e.g., see the solid, dashed
and crosses line in Fig.~\ref{fig:thresh}.
For decreasing $\tau_{snw}/\tau_{c}$, however, both $J_{T1}$
and $J_{T2}$ approach $J_{T}$ (dotted and dash-dotted line), so the values
of $r$ and $d$ decrease. If the time that the electrons spend 
in the WL is not much shorter than $\tau_{snw}$, their spin polarization 
will be largely erased before
capture by the QDs. The typical times, $\tau_c\sim$ 1 to 10 ps,
 make this scenario unlikely. 

The influence of escape time $\tau_{e}$ on the thresholds
is modest. Keeping 
the ratio $\delta=\tau_{c}/\tau_{e}$ fixed, and increasing $\tau
_{c},$ we find similar shifts of $J_{T1}$ and $J_{T2}$ to slightly
higher values. By changing $\delta$ from zero to 1.25 
(zero to high-temperature limit\cite{Summers2007:JAP}),
the spin-filtering region decreases from
$d\!\left(\tau_{c}=10~\mathrm{ps}\right)=1.34$ to 
$d\!\left(\tau_{c}=10~\mathrm{ps}\right)=1.32$, 
with similar changes in the $0.1\ \mathrm{ps}<\tau_{c}<200\ \mathrm{ps}$ 
interval (the other parameters retaining the initial values).

It is interesting to consider the influence of the optical 
matrix element of the lasing transition, $M$.
Increasing $\left|M\right|^2$ results in a
proportional increase of both $g$ and $b_q$,\cite{Chuang:2009}
 representing gain and 
radiative losses, respectively.
The increase of the losses prevails, so that all the thresholds
rise. The value of $J_T\left(\tau_c=0\right)$ is an example:
in Eq.~(\ref{eq:J0}) $f_0$ decreases with increasing
$g$ (increasing $\left|M\right|^2$) to the minimum 1/2, 
but $b_q$ grows indefinitely.
Figure~\ref{fig:thresh} shows the corresponding
change of $J_{T1}$ and $J_{T2}$ on the example of a three-fold
increase of $\left|M\right|^2$.
The matrix element modifies $J_{T1,2}$ 
to a different extent than $J_{T}$. 
With growing $\left|M\right|^2$, the threshold  $J_{T}$ rises faster,
which results in a higher $r$ and $d$.
For example, setting $\tau_c=10$~ps and using the initial parameters,
except for $\tau_{snw}=100$~ps (appropriate for room temperature
\cite{Jarasiunas2004:SST,*Schreiber2007:PSSB}),
we find $r=0.13$, $d=0.52$. These values increase
to $r=0.22$, $d=0.75$ for $g\tau_\mathrm{ph}=8$ and 
$b_q\tau_\mathrm{ph}=0.04$ (a four-fold increase of $\left|M\right|^2$).
We find a similar improvement of $r$ with increasing
photon lifetime $\tau_\mathrm{ph}$. The above value of $r=0.13$ rises
to 0.22, when $\tau_\mathrm{ph}$ changes from 1 to 4 ps.
Thus, the detrimental effect of spin relaxation in WL can be mitigated 
by modifying laser parameters not related to spin.

Apart from increasing the thresholds, the limited supply of carriers to the
lasing transition causes output saturation. This can be understood
by looking at the regime of high injection. High $J_n$
drives the WL occupancies close to saturation,
$f_{wn\pm}=f_{wp}=1$, because
the finite $\tau_c$ limits carrier relaxation to QDs.
In this regime, the capture rates approach
their maxima, $J_L$ [see Eq.~(\ref{eq:JT})], so that the injection into QDs
no longer grows with $J_n$.
 The asymptotic value of photon occupancy
\begin{equation}
f_{\mathrm{S}}^{\max}\equiv
\left[ f_{\mathrm{S}+}\left(  J_{n}\rightarrow\infty\right)%
     +f_{\mathrm{S}-}\left(%
J_{n}\rightarrow\infty\right)\right]\!/2  <\infty,
\label{eq:fmax}
\end{equation}
is independent of $P_{Jn}$.\cite{Note1}
We obtain $f_\mathrm{S}^{\max}\sim1/\tau_{c}$ for $J_L\gg J_0$ .
 Interestingly,
$f_{\mathrm{S}+}\left( J_{n}\rightarrow\infty\right)=
f_{\mathrm{S}-}\left( J_{n}\rightarrow\infty\right)$, 
so that the circular polarization of laser light, 
$P_\mathrm{S}\equiv \left( f_{\mathrm{S}+} - f_{\mathrm{S}-} \right)/%
\left( f_{\mathrm{S}+} + f_{\mathrm{S}-} \right)$,
is zero for high injection, in contrast to the QW model, where
$P_\mathrm{S}\rightarrow-P_{Jn}$.\cite{Gothgen2008:APL} 
This can be explained as follows.
In a QW-laser model with no gain compression term, levels participating
in the laser action are assumed to be replenished instantaneously
(a characteristic relaxation time is $\sim 1\ \mathrm{ps}$, 
Ref.~\onlinecite{Carroll:1998}).
The capture process to the discrete, widely spaced QD
levels is slower,\cite{Schneider2001:PRB} and must be treated explicitly
in a realistic QD model.
As noted above, this leads to $f_{wn\pm}\lesssim 1$ at sufficiently 
high $J_n$, so
the electrons captured into the QDs are spin-unpolarized and consequently
$P_S\rightarrow0$.

Finally, we note that the limited capture rate is not the only difference
between QW- and QD-based lasers. Since $f_{q\alpha\pm}\leq1$,
Eq.~(\ref{eq:f0}) imposes a lower limit on the gain required for lasing: 
$g\tau_\mathrm{ph}\geq1$ for any $P_{Jn},\ \tau_c$ and the other parameters. 
The QW model of Sec.~\ref{Sec.Intro} predicts no such limit.
A more restrictive condition must be satisfied to maintain the full
threshold reduction: $g\tau_\mathrm{ph}\geq1+\left\vert P_{Jn}\right\vert
$ in the $\tau_c\rightarrow0$ limit. Decreasing $g\tau_\mathrm{ph}$ below
$1+\left\vert P_{Jn}\right\vert $ results in a decrease of $r$, which 
vanishes completely, when $g\tau_\mathrm{ph}\rightarrow1$.
These effects are
a direct consequence of the limited density of states at the lasing transition,
a limitation that can be neglected in QW-based lasers operating at low powers.
We also note that the upper bound $f_{\gamma\alpha}\le 1$ must be enforced
by including Pauli-blocking terms, otherwise the REs lead to incorrect 
results, also for $P_{Jn}=0$. For example, 
if the $1-f_q$ term in Eq.~(\ref{eq:C}) is omitted ($f_q$ is any of
the equal QD occupancies), 
the REs allow for the unphysical $f_q>1$,
so that $J_T$ is always reached, even when $g\tau_{\rm ph}<1$. For  
$g\tau_{\rm ph}\gtrsim1$ and $\tau_c\sim 200$ ps, 
the omission of $1-f_q$ 
leads to relative errors of $J_T$ as high as 30\%.
\section{Conclusions}
In this work we have developed a transparent rate-equation approach,
which has allowed for analytical results. Using this formalism,
we have elucidated various trends in operation of QD spin-lasers, 
comparing them to their relatively well-known 
QW-based counterparts.
In particular, we have studied the consequences of finite capture
rate by QD-confined levels, which participate in the lasing transition.
To fully preserve the threshold reduction and the "spin-filtering" effects 
resulting from spin injection,
the capture time has to be much shorter than the spin relaxation time
in the wetting layer.
Nevertheless, we have found that, when the spin relaxation lowers
the electron spin polarization appreciably, 
the threshold reduction and the "spin-filtering" window can 
be partially restored by modifying some spin-independent laser parameters.
Another consequence of the finite 
capture rate is saturation of stimulated emission as a function of injection.
Furthermore, QD- and QW-based lasers have qualitatively different densities of
the initial and final states of lasing transitions. 
The threshold reduction in QD lasers may be hindered
by the small density of QD states, if the the gain $g$ or the photon cavity 
lifetime are too small.

To take full advantage of the potential of electrical spin injection in 
QD spin-lasers, it
is important to further improve their magnetic contacts (injectors). 
The current maximum temperature of 200 K for electrically injected spin-lasers
 using MnAs injector,\cite{Basu2008:APL} will likely be soon improved,
 since the same spin injector material was recently demonstrated to operate
 at room temperature.\cite{Fraser2010:APL} Fe Schottky contacts have 
also been used to inject
spins in (In,Ga)As QDs at room temperature.\cite{Li2005:APL} 
For the surface-emitting spin lasers, magnetic injectors with out-of-plane 
remanent
 magnetization would be desirable. \cite{Hovel2008b:APL} In such a geometry,
 the spin-laser operation is possible without the need to apply an external
 magnetic field, since the optical selection rules lead to circularly
 polarized light.\cite{Zutic2004:RMP} Encouraging results have been reported
 recently for spin light-emitting diodes utilizing MgO tunnel contacts,%
\cite{Hovel2008b:APL}
 which provide a very efficient room-temperature spin injection.%
\cite{Jiang2005:PRL}

\section{Acknowledgements}
This work was supported by the U.S. ONR, AFOSR-DCT, NSF-ECCS CAREER, and 
DOE-BES.

%
\end{document}